# Vertical transverse transport induced by hidden in-plane Berry curvature in two dimensions

Kyoung-Whan Kim[1†], Hogyun Jeong[2†], Jeongwoo Kim[3*] & Hosub Jin[2*]

[1]*Center for Spintronics, Korea Institute of Science and Technology, Seoul 02792, Korea*

[2]*Department of Physics, Ulsan National Institute of Science and Technology, Ulsan 44919, Korea*

[3]*Department of Physics, Incheon National University, Incheon 22012, Korea*

## Abstract

**The discovery of Berry curvature (BC) has spurred a tremendous surge of research into various quantum phenomena such as the anomalous transport of electrons and the topological phases of matter. In two-dimensional crystalline systems, the conventional definition of the BC lacks the in-plane components and thus it cannot explain the transverse transport along the plane-normal direction. Here, we modify the BC to newly provide in-plane components in two dimensions, giving rise to the vertical Hall effects that describe out-of-plane transports in response to in-plane perturbations and their Onsager reciprocity. Our first-principles calculations show that a large in-plane BC can appear even in an atomic-thick $GdAg_2$ monolayer, and a hexagonal $BiAg_2$ monolayer can host a large BC dipole known to vanish in the conventional BC. The quantum transports driven by the hitherto-hidden BC will become more significant in recently emerging two-dimensional platforms, including van der Waals heterostructures.**

[†]K.-W. K. and H. Jeong contributed equally to this work.
*Corresponding authors. Email: kjwlou@inu.ac.kr, hsjin@unist.ac.kr

**Introduction**

The anomalous transport of electrons and the topological phases of matter originate from the quantum geometric phase of Bloch states, which is described by the Berry curvature (BC) [1-3]. In three-dimensional periodic systems, the crystal momentum **k** is a good quantum number and the reduced Hamiltonian $H(\mathbf{k})$ is acquired by the Bloch theorem. The eigenstates and the corresponding eigenvalues are denoted by $|n(\mathbf{k})\rangle$ and $E_n(\mathbf{k})$. The BC of the *n*-th band has the conventional form of [4,5]

$$\mathbf{\Omega}_n^{\mathrm{3D}}(\mathbf{k}) = i \sum_{m \neq n} \frac{\langle n|\nabla_{\mathbf{k}} H|m\rangle \times \langle m|\nabla_{\mathbf{k}} H|n\rangle}{(E_n - E_m)^2}. \tag{1}$$

Hereafter, unless specified, we express $|n(\mathbf{k})\rangle$ and $E_n(\mathbf{k})$ by $|n\rangle$ and $E_n$ for simplicity. A nonvanishing BC appears in symmetry-broken environments [5]. For a system with broken time-reversal symmetry, the integration of $\mathbf{\Omega}(\mathbf{k})$ over the momentum space links to an intrinsic anomalous Hall effect [3,6]; for a noncentrosymmetric system, the integration of $\partial_{k_i}\mathbf{\Omega}(\mathbf{k})$, referred to as the BC dipole, gives rise to the quantum nonlinear Hall effect [7-9] and the photogalvanic effect [10-13].

By contrast, for two-dimensional systems located in the *xy* plane, $k_z$ is not a good quantum number and only the *z* component of the BC is defined by Eq. (1). Through the dimensional crossover from three to two dimensions (Fig. 1), one can gain insight into the incompleteness of the existing definition of the BC. The left panels of Fig. 1 show three-dimensional materials with broken time-reversal symmetry and broken inversion symmetry, respectively. When a ferromagnetic moment is present along the *x* direction, the net flux of the *x*-component BC induces an anomalous Hall effect. As illustrated in Fig. 1(a), for example, an electric field along the $z$ direction ($E_z$) generates an electrical current along the $y$ direction ($J_y$) [3]. Similarly, when electric polarization is present along the $z$ direction, a helical structure of BC around the polarization ($\mathbf{\Omega} \sim \hat{\mathbf{z}} \times \mathbf{k}$) carries a nonzero BC dipole, leading to a nonlinear Hall current ($J_z$) in response to an oscillating electric field ($E_{\omega,x}$), as shown in Fig. 1(b) [11,14]. All of these transverse transport phenomena are engendered by the well-defined $x$, $y$ components of the BC in three dimensions. We now reduce to the lower dimension by gradually decreasing the thickness of the system (Fig. 1, right panels). Because the symmetry breaking remains the same, one can speculate that the aforementioned anomalous responses would be retained during the dimensional reduction. In a recent experiment [15], the giant vertical nonlinear Hall effect has been reported upon varying the thickness of $WTe_2$ and $MoTe_2$ films. The conventional BC in two dimensions, however, cannot describe these responses because Eq. (1) does not include the in-plane components. Therefore, it is necessary to generalize the BC formula to provide all three components for a complete description and applications of anomalous transport in two dimensions.

In this work, we generalize the BC in two dimensions and predict the generation of a large vertical Hall current in response to an in-plane electric field. The reformulated BC has in-plane components that are absent in the conventional BC. Using perturbation theory, we derive that the in-plane BC governs the vertical transverse transports illustrated in the right panels of Fig. 1. Our density functional theory (DFT) calculations reveal that the in-plane BC can dominate over the conventional out-of-plane component even in the extreme limit of atomic-thick two-dimensional systems in which the vertical motion is suppressed. We consider two representative examples of inversion-broken and time-reversal-broken systems: $BiAg_2$ and $GdAg_2$ monolayers. Against the general belief that the BC dipole vanishes in two-dimensional hexagonal systems [7], a large dipole component of BC survives in a BiAg$_2$ monolayer based on our formalism. We also performed the time-dependent DFT calculations, explicitly demonstrating the vertical Hall currents induced by the in-plane BC.

**Generalization of BC in two dimensions**

We present the following generalization of the BC by applying a position operator formalism [16-18]. In a periodic system, the physical meaning of $\nabla_{\mathbf{k}} H$ in Eq. (1) is the velocity operator multiplied by $\hbar$. Therefore, $\langle n|\partial_{k_z} H|m\rangle$ can be replaced with $\langle n|i[H,z]|m\rangle = i(E_n - E_m)\langle n|z|m\rangle$ for $n \neq m$ in a system confined along the $z$ direction. Notably, $\langle n|z|m\rangle$ is well-defined in a finite system, enabling the in-plane BC components to be reformulated. The modified version of Eq. (1) for two-dimensional systems is then

$$\boldsymbol{\Omega}_n^{2D}(\mathbf{k}) = i \sum_{m \neq n} \frac{\langle n|\nabla_{\mathbf{k}} H|m\rangle \times \langle m|\nabla_{\mathbf{k}} H|n\rangle}{(E_n - E_m)^2} + 2\mathrm{Re} \sum_{m \neq n} \frac{\langle n|\nabla_{\mathbf{k}} H|m\rangle \times \langle m|z\hat{\mathbf{z}}|n\rangle}{E_n - E_m}. \quad (2)$$

The first term is the conventional out-of-plane component, and the second term gives in-plane BC, which has not yet been considered. Unlike $\boldsymbol{\Omega}^{3D}$ in Eq. (1), the two-dimensional $H(\mathbf{k})$ is not sufficient to express $\boldsymbol{\Omega}^{2D}$, requiring additional information about the *hidden* degree of freedom along the $z$ direction. Note that the second term in Eq. (2) is not strictly a *curvature* in the conventional sense, which originates from the holonomy of the electronic states in $\mathbf{k}$ space. But in this work we call it a generalized version of a curvature because it becomes equivalent to the conventional curvature when the periodicity along the $z$ direction is restored. We also note that similar generalizations are possible for one and zero dimensions [19]. We hereafter omit the superscript "2D" unless specified.

The reformulated BC indeed describes the anomalous responses illustrated in the right panels in Fig. 1. In the presence of a constant electric field ($E$) along the $z$ direction, the perturbed eigenstate is written as $\widetilde{|n\rangle} = |n\rangle - \sum_{m \neq n} |m\rangle \langle m|eEz|n\rangle / (E_n - E_m)$ in the length gauge. The transverse in-plane current is then given by the in-plane components of the reformulated BC:

$$\mathbf{J} = -\frac{e}{\hbar} \sum_{\mathbf{k},n} f_{\mathbf{k}} \langle \widetilde{n}|\nabla_{\mathbf{k}} H |\widetilde{n}\rangle = \frac{e^2}{\hbar} E \hat{\mathbf{z}} \times \sum_{\mathbf{k},n} f_{\mathbf{k}} \mathbf{\Omega}_n(\mathbf{k}), \quad (3)$$

where $f_{\mathbf{k}}$ is the electron distribution function. In two dimensions, Eqs. (2) and (3) account for a new type of anomalous Hall transport in response to a vertical electric field [Fig. 1(a), right panel]. Also, the Onsager reciprocity indicates that an in-plane electric field can induce the $z$-directional shift of an electron so that a charge current is pumped into an adjacent metal electrode stacked in the vertical direction. In the absence of the inversion symmetry, a semiclassical theory [7] shows that the nonequilibrium distribution function perturbed by an external electric field, $\delta f_{\mathbf{k}} \propto \mathbf{E} \cdot \nabla_{\mathbf{k}} f_{\mathbf{k}}$, may generate the nonlinear Hall and photogalvanic currents [Fig. 1(b), right panel], whose magnitudes are both proportional to the BC dipole.

**Helical in-plane BC in a polar BiAg$_2$ monolayer**

Here we investigate the generalized BC in an atomic-thick BiAg$_2$ monolayer, in which the out-of-plane polar displacement ($\Delta z$) of Bi atoms [Fig. 2(a)] [20] induces substantial in-plane BC and its dipole. Figure 2(b) is the electronic band structure calculated from the first-principles calculations [19] and shows a Rashba-like spin splitting induced by the inversion asymmetry. As illustrated in Fig. 1(b), the polar displacement is expected to display a helical structure of the in-plane BC and the corresponding out-of-plane BC dipole. In Fig. 2(c), a clockwise helical BC vector is drawn in the momentum space with the chemical potential $\mu$ set to 0.55 eV (green horizontal lines). Figure 2(d) shows the out-of-plane BC dipole density, $d_z(\mathbf{k}) = (1/2)[\nabla_\mathbf{k} \times \mathbf{\Omega}(\mathbf{k})]_z$, arising from the helical in-plane BC, whose maximum size reaches 500 Å$^3$. By integrating the BC dipole density over the first Brillouin zone, we plot the BC dipole as a function of $\mu$ in Fig. 2(e). According to Fig. 2(f), the helical in-plane BC is a direct consequence of the polar displacement.

The resulting BC dipole implies the significance of our generalized formalism. According to the conventional BC formula, the BC dipole vanishes in two-dimensional crystals belonging to $C_{3v}$ symmetry class [7], and hence it has been widely believed that the nonlinear Hall effect is absent in two-dimensional hexagonal crystals. On the contrary, our calculations exhibit a *hidden* BC dipole in a BiAg$_2$ monolayer, which is also in the $C_{3v}$ symmetry class. Furthermore, the maximum size of the BC dipole density is more than two orders of magnitude larger than that arising from the conventional out-of-plane BC in III–V semiconductor [110] quantum wells (~1 Å$^3$) [12]. The resultant BC dipole is as large as −0.4 Å, which is comparable [19] to the enormously enhanced value at the topological phase transition in BiTeI [14]. It is also comparable to the large BC dipole (induced by the out-of-plane BC) of WTe$_2$ multilayers [8,9,13] and SnTe monolayers [21].

To investigate the microscopic mechanism of the in-plane BC in a BiAg$_2$ monolayer, we constructed a four-band model based on *p* orbitals in Bi atoms and *s* orbitals in Ag atoms. As mentioned above [after Eq. (2)], unlike the conventional BC, the in-plane BC cannot be calculated only by the Hamiltonian and its eigenstates, but additional information on matrix elements of the out-of-plane position operator *z* is necessary. By carefully taking into account them, we derive [19]

$$\mathbf{\Omega}_{\text{in}}^{p_z} \propto \Delta z \hat{\mathbf{z}} \times \mathbf{k}, \tag{4}$$

up to first order in *sp* hybridization and the polar displacement. Here $\mathbf{\Omega}_{\text{in}}^{p_z}$ is the in-plane BC for the $p_z$ band and $\Delta z$ is the polar displacement of Bi atoms [Fig. 2(a)]. The in-plane BC forms a helical BC structure induced by the inversion breaking polar displacement as we depict in Fig. 1(b). The out-of-plane BC dipole density from the helical BC texture is then given by $\frac{1}{2}\nabla_{\mathbf{k}} \times \mathbf{\Omega}_{\text{in}}^{p_z} \propto \Delta z \hat{\mathbf{z}}$. The in-plane BCs for the other bands show similar behaviors, as implied by the symmetry. Recalling that the matrix elements of the position operator *z* are essential for our calculation, the finite spatial extension of electron clouds along the surface normal direction and inter-orbital mixing triggered by the polar displacement plays a crucial role in providing the in-plane BC.

### In-plane BC flux in a ferromagnetic GdAg$_2$ monolayer

Now we consider the second system, a GdAg$_2$ monolayer, where the in-plane ferromagnetic moments of Gd atoms break the time-reversal symmetry [22,23]. The electronic band structure corresponding to the Gd moments aligned along the *x* direction [Fig. 3(a)] is shown in Fig. 3(b). As illustrated in Fig. 1(a), the in-plane ferromagnetic moment yields the *yz* component of the anomalous Hall conductivity ($\sigma_{yz}$) given by the integration of the *x* component BC ($\Omega_x$) multiplied by $e^2/h$ [Eq. (3)]. The calculated $\sigma_{yz}$ reaches 0.62 $e^2/h$ in the vicinity of the Fermi level [Fig. 3(c)], which is comparable to the quantized Hall conductivity of Chern insulators [24,25]. The substantial size of $\sigma_{yz}$ in the GdAg$_2$ monolayer implies that, even in atomically thin films, the vertical Hall response cannot be neglected. The newly unveiled $\sigma_{yz}$ can provide a proper interpretation of some measurement schemes whose microscopic formulations in two dimensions remain elusive, such as the longitudinal magneto-optic Kerr effect of in-plane two-dimensional ferromagnets [26,27].

Figures 3(d)-3(f) show maps of the three components of the BC when the chemical potential is tuned slightly above the Fermi level [yellow line in Fig. 3(b)]. In Fig. 3(d), the maximum magnitude of $\Omega_x$ (283 Å$^2$) emerges along the Γ–K line from the anticrossing between $m_z = \pm 1$ and $m_z = \pm 2$ bands composed of $d_{yz}/d_{zx}$ and $d_{xy}/d_{x^2-y^2}$ orbitals, respectively. The integration of $\Omega_y$ vanishes because of the $yz$ mirror plane perpendicular to the ferromagnetic moment [Fig. 3(e)], and $\Omega_z$ in Fig. 3(f) is negligible ($\lesssim 1$ Å$^2$).

In Supplemental Material [19], we construct a model Hamiltonian and the matrix element of the position operator $z$. We calculate the in-plane BC for one of the $m_z = \pm 2$ bands as

$$\mathbf{\Omega}_{\text{in}} \propto \alpha k_x \mathbf{k}, \tag{5}$$

up to first order in spin-orbit coupling parameter α and the inverse of the band splitting energy. Note that $\Omega_x \propto k_x^2$ can give a nonvanishing BC monopole, while $\Omega_y \propto k_x k_y$ cannot. And the in-plane BC distribution of $\Omega_x \propto k_x^2$ and $\Omega_y \propto k_x k_y$ is also consistent with $C_{2x}$ rotation and $M_x$ mirror reflection symmetry of the system. These features are consistent with Figs. 3(d)-3(e). Our analytic theory reveals that, under the in-plane spin polarization, spin-orbit coupling acts as a ladder operator to raise or lower the orbital angular momentum, which intertwines the lateral and vertical movements of electrons and consequently gives rise to the large $\Omega_x$.

**Time-dependent DFT calculations for the vertical Hall effect**

By performing time-dependent DFT calculations [19] for a GdAg$_2$ monolayer, we explicitly demonstrate the anomalous Hall current induced by the in-plane BC shown in Fig. 3(c). Figure 4(a) shows the time-varying electric field obtained by the Fourier transform of the frequency spectrum shown in the inset. By applying the electric field along the *y* direction ($E_y$), we obtain the transverse current along the *z* direction ($J_z$), which is perpendicular to both the electric field (**E**) and the magnetic moment (**M**) [red curve in Fig. 4(b)]. We also observe an oscillatory current profile whose period approximately coincides with that of the electric field. Because of the lack of an additional relaxation process in our simulation, the oscillating current persists even after $t = 2.0$ fs when the external field is negligible. The blue curve in Fig. 4(b) shows the Hall current along the *y* direction ($J_y$) under an electric field along the *z* direction. $J_z$ and $J_y$ are of the same magnitude but with opposite signs, clearly showing the expected Onsager reciprocity $\sigma_{yz} = -\sigma_{zy}$.

In addition, we investigated the out-of-plane current response by varying the direction of the applied electric field within the plane [Fig. 4(c)]. Here, $\theta$ is the angle of the electric field that deviates from the *y* direction [Fig. 4(d)]. The magnitude of $J_z$ is maximum when the electric field is perpendicular to the direction of the ferromagnetic moment (black and purple curves), whereas it completely vanishes when the electric field is parallel to **M** (green curve). Figure 4(e) shows a plot of the normalized Hall current as a function of $\theta$ [19]. Despite the complicated oscillating feature of $J_z$ in time, all data fall along a single cosine curve, confirming that $\mathbf{J}_H \sim \mathbf{E} \times \mathbf{\Omega}$. Therefore, the transverse current $J_z$ is directly determined by the anomalous Hall conductivity $\sigma_{zy}$ in Eq. (3) and it thus is the manifestation of the in-plane BC. The transverse transport phenomena in our simulation indeed represent a new type of anomalous Hall effect in a two-dimensional in-plane ferromagnet whose response function is governed by the in-plane BC.

**Conclusion**

We generalize the BC in two dimensions, unravelling, in particular, the existence of the in-plane components and corresponding responses. The in-plane components capture linear and nonlinear anomalous Hall effects along the surface normal direction, which cannot be described by the conventional BC formalism. The vertical responses originate from the spatial distribution of electrons along the confined direction and are inevitable in real two-dimensional systems. The giant in-plane BC appearing in the atomic-thick limit represents an advancement beyond proof-of-concept and provides a new perspective into quantum transport in low dimensions. Furthermore, in the currently prevailing van der Waals heterostructures [28-31], the generalized BC can be tailored through adjustment of the symmetry in diverse ways, enriching the scope and functionality of two-dimensional systems.

**Acknowledgments**

K.-W. K was supported by the KIST Institutional Program (2E31032) and the National Research Foundation of Korea (NRF) (2020R1C1C1012664, 2019M3F3A1A02071509). J. K. was supported by an NRF grant funded by the Korea government (MSIT) (No. 2020R1F1A1048143). H. Jeong and H. Jin were supported by the NRF under Grant Nos. 2021M3H4A1A03054864, 2019R1A2C1010498 and 2017M3D1A1040833.

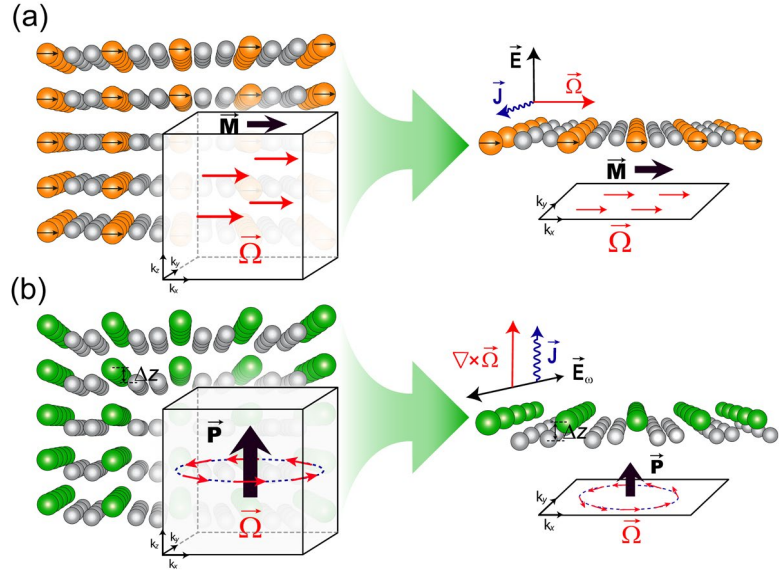

FIG. 1. (a) (left panel) The BC flux is induced by broken time-reversal symmetry in a three-dimensional solid. (right panel) In dimensional crossover preserving the symmetry breaking, in-plane BC appears with the corresponding linear anomalous Hall effect in two dimensions. (b) (left panel) The helical BC texture emerges from broken inversion symmetry in a three-dimensional solid. (right panel) During dimensional reduction, the helical texture of in-plane BC is preserved, resulting in the out-of-plane BC dipole and the corresponding nonlinear anomalous Hall effect in two dimensions.

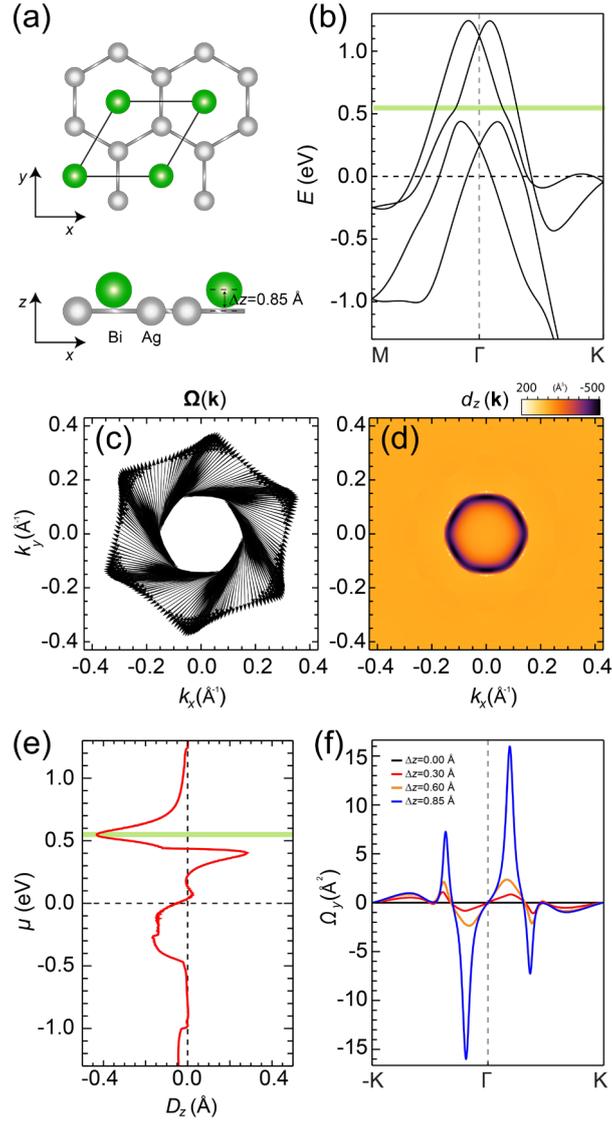

FIG. 2. (a) Atomic structure of the BiAg$_2$ monolayer. The polar displacement of the Bi atoms breaks the inversion symmetry. (b) The electronic structure of the BiAg$_2$ monolayer. (c) The in-plane BC vectors in the momentum space, as calculated at the chemical potential $\mu = 0.55$ eV. (d) The BC dipole density ($d_z$) in the momentum space calculated from the curl of the BC vectors in (c). (e) Plot of BC dipole ($D_z$) (the integration of $d_z$ over the first Brillouin zone) as a function of the chemical potential. (f) $\Omega_y$ calculated along the $k_x$ direction for $k_y = 0$ by varying the polar displacement. The result indicates that the in-plane BC is proportional to the amount of symmetry breaking induced by the polar displacement.

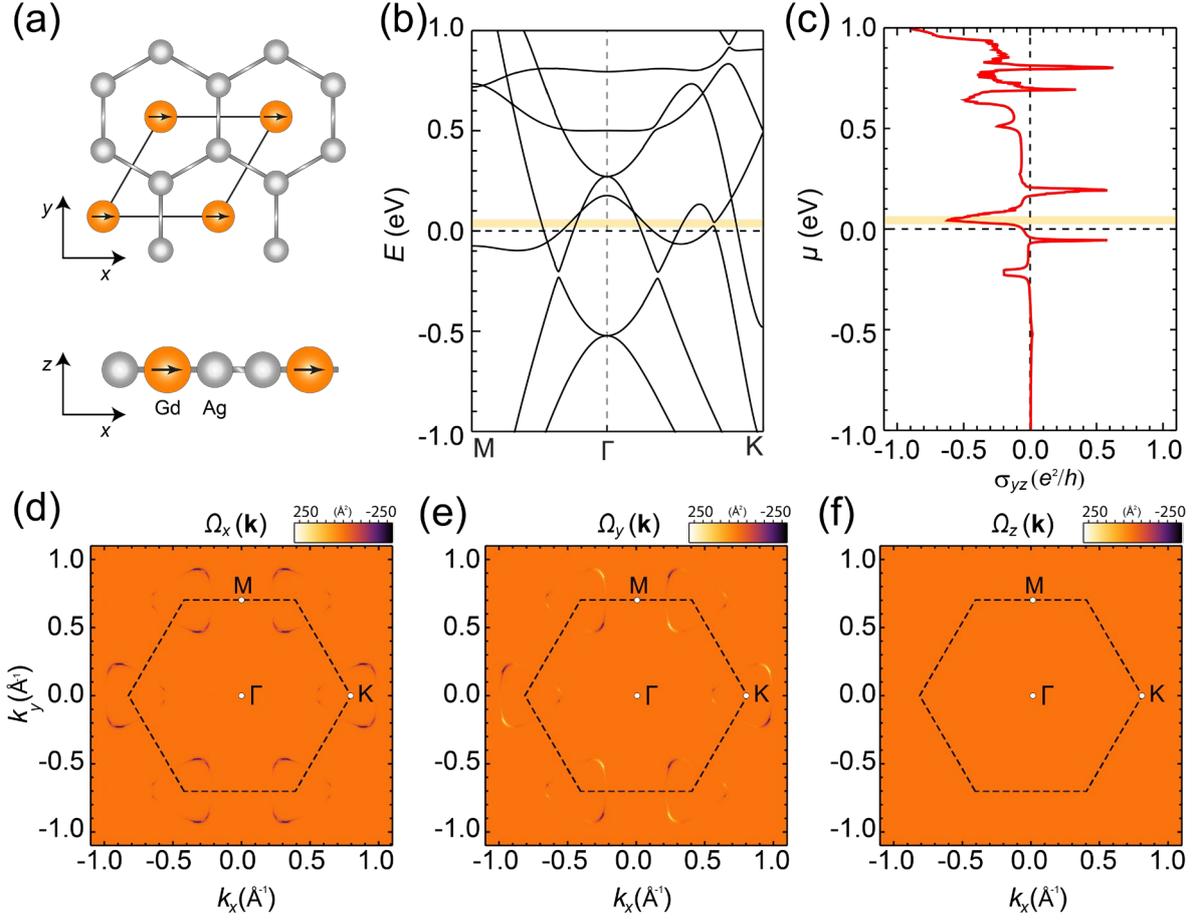

FIG. 3. (a) Atomic structure of the GdAg$_2$ monolayer. The in-plane ferromagnetic moment of the Gd atoms breaks the time-reversal symmetry. (b) The electronic structure of the GdAg$_2$ monolayer. (c) The $yz$ component of the anomalous Hall conductivity (the integration of $\Omega_x$ over the first Brillouin zone) as a function of the chemical potential ($\mu$). (d)-(f) Maps of each component of $\mathbf{\Omega}$ calculated at the chemical potential depicted as yellow lines in (b) and (c). The results are consistent with the $yz$ mirror reflection and inversion symmetry of the system. Whereas $\Omega_x$ is large, the other components are negligible ($\Omega_z$) or vanish when integrated out ($\Omega_y$).

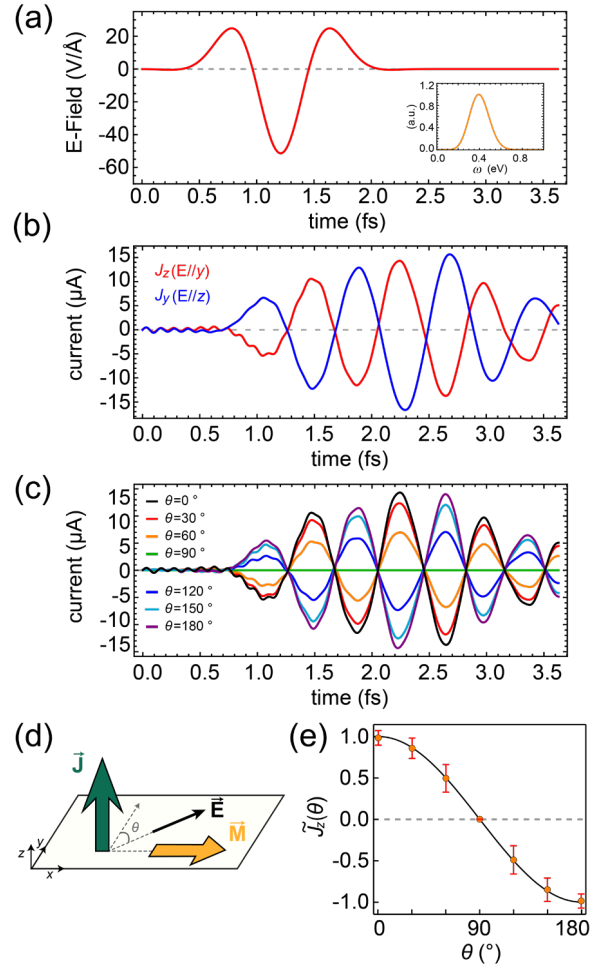

FIG. 4. (a) The applied electric field to the GdAg$_2$ monolayer as a function of time in the time-dependent DFT calculations. (inset) Frequency spectrum of the applied electric field. (b) Time-varying current along the $y$ direction ($J_y$) under the electric field along the $z$ direction (red) and that along the $z$ direction ($J_z$) under the electric field along the $y$ direction (blue). (c) The $J_z$ calculated by rotating the in-plane directions ($\theta$) of the applied electric field. (d) Schematic of the geometry of the simulation. (e) Time-averaged normalized currents as a function of $\theta$. The error bars are determined by the standard deviation over the evaluation times. The solid black line denotes $\cos\theta$.

Supplemental Material for

# Vertical transverse transport induced by hidden in-plane Berry curvature in two dimensions

Kyoung-Whan Kim[1†], Hogyun Jeong[2†], Jeongwoo Kim[3*] & Hosub Jin[2*]

[1]*Center for Spintronics, Korea Institute of Science and Technology, Seoul 02792, Korea*

[2]*Department of Physics, Ulsan National Institute of Science and Technology, Ulsan 44919, Korea*

[3]*Department of Physics, Incheon National University, Incheon 22012, Korea*

## I. Generalized BC in lower dimensions

Our generalization can be easily adopted to suggest the BC in lower dimensions, where the conventional definition [Eq. (1)] can provide *none* of its components. In a one-dimensional system aligned along the $x$ direction, for example, only $k_x$ is the well-defined crystal momentum. We can then formulate $\Omega_y$ and $\Omega_z$ in a similar manner through Eq. (2) by considering the $x$-component velocity operator ($\partial_{k_x} H$) and the position operators ($z$ and $y$). In addition, $\Omega_x$ is given as

$$\Omega_x^{\text{1D}}(\mathbf{k}) = -2\text{Im} \sum_{m \neq n} \langle n|y|m\rangle\langle m|z|n\rangle. \tag{S1}$$

Note that Eq. (S1) is valid even in zero-dimensional systems in which the crystal momentum is not defined. The position operator can be utilized in defining all three components of the BC.

$$\boldsymbol{\Omega}^{\text{0D}} = -\text{Im} \sum_{m \neq n} \langle n|\mathbf{r}|m\rangle \times \langle m|\mathbf{r}|n\rangle. \tag{S2}$$

where $\mathbf{r} = (x, y, z)$.

## II. Details for first-principles calculations

For the DFT calculations, we used the full-potential linearized augmented-plane-wave (FP-LAPW) method, as implemented in the Elk code [32]. The Perdew–Burke–Ernzerhof-type generalized gradient approximation was used to treat the exchange-correlation interaction among electrons [33]. The muffin-tin radii ($R_{\text{MT}}$) were chosen as 2.8, 2.8, and 2.6 a.u. for Bi, Gd, and Ag, respectively. The product of the muffin-tin radius and the maximum reciprocal lattice vector ($R_{\text{MT}} \times |\boldsymbol{G} + \boldsymbol{k}|_{\text{max}}$) was set to 8. For the electronic structure calculations, we used a $32 \times 32 \times 1$ $k$-grid to sample the full Brillouin zone. To precisely simulate the isolated two-dimensional materials, the supercell geometries were separated by 30 Å of vacuum in the surface normal direction. For the GdAg$_2$ monolayer, we used the LDA+$U$ scheme with the fully localized-limit double-counting correction to appropriately describe the correlation effect of a Gd $f$ orbital ($U = 6.5$ eV and $J = 0.5$ eV) [34].

Using the Kohn–Sham eigenstates $\psi_{n\mathbf{k}}$ obtained from our DFT calculations and the generalized BC formalism of Eq. (2), we evaluated the in-plane BC in BiAg$_2$ and GdAg$_2$ monolayers. The anomalous Hall conductivity is expressed as

$$\sigma_{\alpha\beta} = -\frac{e^2}{\hbar} \int \frac{d^2k}{(2\pi)^2} \sum_{n,m\neq n} f_{n\mathbf{k}} \frac{-2\mathrm{Im}\langle\psi_{n\mathbf{k}}|v_\alpha|\psi_{m\mathbf{k}}\rangle\langle\psi_{m\mathbf{k}}|v_\beta|\psi_{n\mathbf{k}}\rangle}{(\omega_{m\mathbf{k}} - \omega_{n\mathbf{k}})^2 + \eta}, \tag{S3}$$

where $\alpha, \beta$ are cartesian coordinates, energy eigenvalues $E_{n\mathbf{k}} = \hbar\omega_{n\mathbf{k}}$, and $\eta$ is a positive infinitesimal. We inserted $\eta = 10^{-6}$ for the numerical stability. Along the $z$ direction, the matrix element of the velocity operator is written as $\langle\psi_{n\mathbf{k}}|v_z|\psi_{m\mathbf{k}}\rangle = \langle\psi_{n\mathbf{k}}|(i/\hbar)[\mathcal{H}_{KS}, z]|\psi_{m\mathbf{k}}\rangle = (i/\hbar)(E_{n\mathbf{k}} - E_{m\mathbf{k}})\langle\psi_{n\mathbf{k}}|z|\psi_{m\mathbf{k}}\rangle$. Exponential decay of $\psi_{n\mathbf{k}}$ into the vacuum allows explicit computation of matrix element of the position-operator. Alternatively, the commutator $v_z = (i/\hbar)[\mathcal{H}_{KS}, z]$ can be directly calculated by inserting the explicit expression of $\mathcal{H}_{KS}$, which results in two terms, i.e., a real-space derivative of wave functions and $[\mathbf{L}\cdot\mathbf{S}, z]$ derived from the spin-orbit coupling. The BC dipole was numerically estimated as [35-38]

$$\mathcal{D}_{\alpha\beta} = \int \frac{d^2k}{(2\pi)^2} \sum_n f_{n\mathbf{k}} \frac{\partial \Omega_{n,\beta}(\mathbf{k})}{\partial k_\alpha}, \tag{S4}$$

where $\alpha, \beta$ are cartesian coordinates and $\Omega_{n,\beta}(\mathbf{k})$ denotes a vector component of $n$-th band BC. The out-of-plane BC dipole is defined as $D_z = (\mathcal{D}_{xy} - \mathcal{D}_{yx})/2$. For computing the BC distribution and its dipole, we used a $300 \times 300 \times 1$ $k$-grid in the Brillouin zone, and unoccupied states were included up to 0.5 Hartree energy above the Fermi level.

## III. Comparison of the BC dipole with three-dimensional materials

BiTeI is a three-dimensional polar insulator that exhibits a large enhancement of its BC dipole by forming a tilted Weyl semi-metallic phase in the vicinity of the topological phase-transition point. Because BiTeI is a three-dimensional material, a direct comparison of its BC dipole with that calculated for a BiAg$_2$ monolayer requires a factor conversion. To this end, we take the effective thickness of a BiAg$_2$ monolayer by summing the atomic radii of Bi and Ag atoms and the polar displacement between them [see Fig. 2(a)], which is sufficient for an order-of-magnitude estimation. Their numerical values are 2.3 Å, 1.7 Å, and 0.85 Å, respectively, which sum to 4.85 Å. Dividing the BC dipole for a BiAg$_2$ monolayer (−0.4 Å) by the effective thickness gives a converted BC dipole on the order of 0.1, which is comparable to the BC dipole in BiTeI at the topological phase transition [14].

## IV. Model for the in-plane BC and its dipole in a BiAg₂ monolayer

The basis set of the model Hamiltonian is acquired from the orbital characters of each Bloch state in the absence of spin-orbit coupling. In Figs. S1(a)-S1(b), the lower two bands mainly show $p_z$ and $s$ characters, and the upper two bands are composed of the radial $p$ orbital ($|p_r\rangle$) and tangential $p$ orbital ($|p_t\rangle$). In an isotropic model, these planar $p$ orbitals can be expressed as $|p_r\rangle = \cos\theta_{\mathbf{k}} |p_x\rangle + \sin\theta_{\mathbf{k}} |p_y\rangle$ and $|p_t\rangle = -\sin\theta_{\mathbf{k}} |p_x\rangle + \cos\theta_{\mathbf{k}} |p_y\rangle$, where $\mathbf{k} = k(\cos\theta_{\mathbf{k}}, \sin\theta_{\mathbf{k}})$ [see Fig. S1(c)].

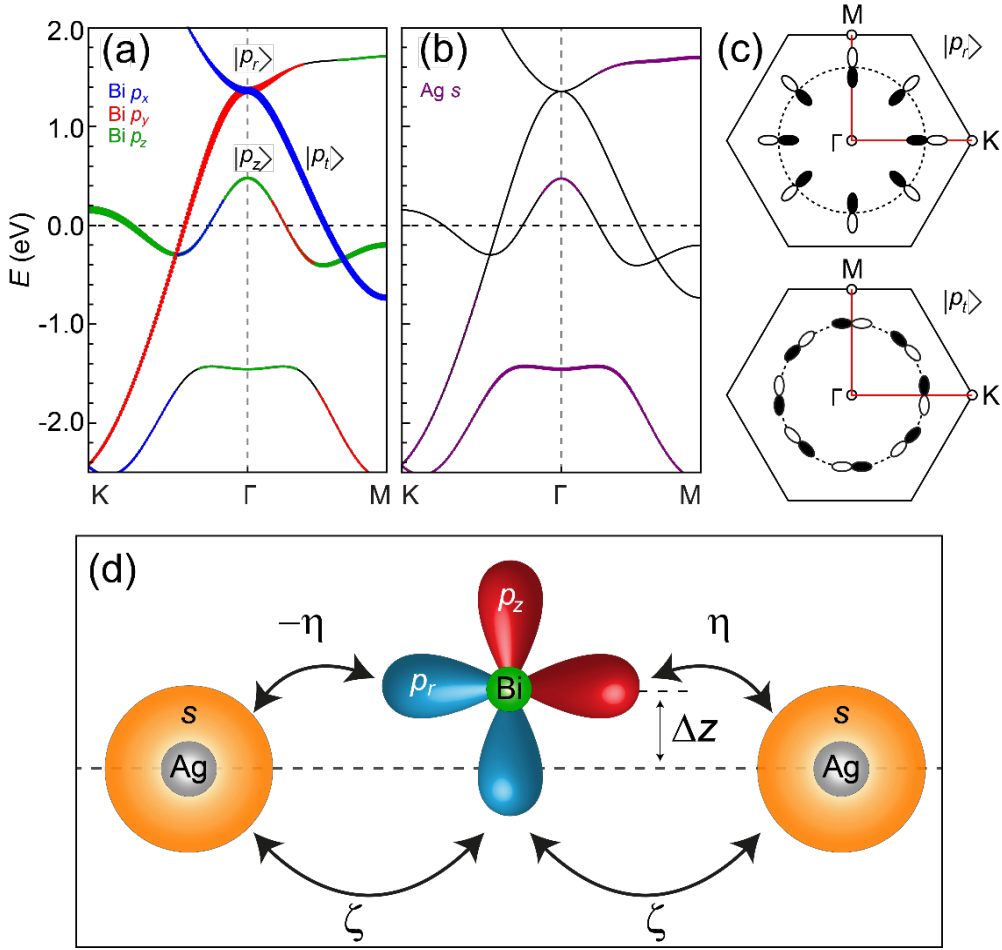

FIG. S1. (a)-(b) The orbital characteristics of a BiAg₂ monolayer in the absence of spin-orbit coupling. The weights of $p_x$, $p_y$, $p_z$, and $s$ are denoted by blue, red, green, and purple dots, respectively. (c) The schematics of orbital textures in the $\mathbf{k}$ space for the radial ($p_r$) and tangential ($p_t$) bands, near $E = 1.4$ eV. (d) Illustration of non-zero position matrix elements between $s$ and $p$ orbitals.

Within the $(|p_r\rangle_{Bi}, |p_t\rangle_{Bi}, |p_z\rangle_{Bi}, |s\rangle_{Ag})$ basis set, the total Hamiltonian is written as the sum of the kinetic Hamiltonian ($H_0$) and the symmetry breaking Hamiltonian ($H_1$) induced by the polar displacement. The kinetic Hamiltonian near the $\Gamma$ point reads

$$H_0 = \begin{pmatrix} E_p + a_1 k^2 & 0 & 0 & -i\gamma k \\ 0 & E_p + a_2 k^2 & 0 & 0 \\ 0 & 0 & a_2 k^2 & 0 \\ i\gamma k & 0 & 0 & E_s + bk^2 \end{pmatrix}, \tag{S5}$$

where $E_p$ and $E_s$ are the band-edge energies of the planar $p$ and $s$ orbitals, respectively, and the energy of the $p_z$ band at the $\Gamma$ point set to be zero. $a_1$ ($a_2$) corresponds to the $\sigma$ ($\pi$) hopping integral, $b$ corresponds to the $s$ orbital hopping integral, and $\gamma$ is the $sp$ hybridization. $H_1$ is given by an additional hopping from $s$ to $p_z$ allowed by the polar displacement of Bi atoms:

$$H_1 = \begin{pmatrix} 0 & 0 & 0 & 0 \\ 0 & 0 & 0 & 0 \\ 0 & 0 & 0 & \xi \\ 0 & 0 & \xi & 0 \end{pmatrix}, \tag{S6}$$

where $\xi$ is proportional to the polar displacement $\Delta z$.

To diagonalize the total Hamiltonian $H = H_0 + H_1$, we treat $\Delta z$ (thus $\xi$) and $\gamma$ pertubatively and keep their first order contributions (not neglecting the cross contributions). The diagonalization includes two procedures: i) integrating out the $s$ orbital and obtaining the orbital Rashba effect and ii) diagonalizing the orbital Rashba contribution.

The $s$ orbital can be integrated out by the following Schrieffer–Wolff transformation. We introduce a unitary transform defined by

$$U_1 = e^{-S_1}, \tag{S7a}$$

$$S_1 = \begin{pmatrix} 0 & 0 & 0 & \frac{i\gamma k}{\Delta E_{sp} + (b - a_1)k^2} \\ 0 & 0 & 0 & 0 \\ 0 & 0 & 0 & -\frac{\xi}{E_s + (b - a_2)k^2} \\ \frac{i\gamma k}{\Delta E_{sp} + (b - a_1)k^2} & 0 & \frac{\xi}{E_s + (b - a_2)k^2} & 0 \end{pmatrix}, \tag{S7b}$$

where $\Delta E_{sp} = E_s - E_p$. Then, the transformed Hamiltonian is now block-diagonal with respect to the $p$ and $s$ orbital blocks.

$$U_1^\dagger H U_1 = \begin{pmatrix} E_p + a_1 k^2 & 0 & ik\alpha_L(k) & 0 \\ 0 & E_p + a_2 k^2 & 0 & 0 \\ -ik\alpha_L(k) & 0 & a_2 k^2 & 0 \\ 0 & 0 & 0 & E_s + bk^2 \end{pmatrix}, \quad \text{(S7c)}$$

$$\alpha_L(k) = \frac{\gamma\xi}{E_s + (b-a_2)k^2}\left\{1 - \frac{E_p + (a_1 - a_2)k^2}{2[\Delta E_{sp} + (a_1 - b)k^2]}\right\}. \quad \text{(S7d)}$$

The combined action of the *sp* hybridization and the polar displacement results in the complex mixing of the $p_r$ state and the $p_z$ state, i.e., the orbital angular momentum along the tangential direction $L_t \equiv i(|p_r\rangle\langle p_z| - |p_z\rangle\langle p_r|)$. The off-diagonal elements of the Hamiltonian is $\sim \alpha_L k L_t = \alpha_L \hat{\mathbf{z}} \cdot (\mathbf{k} \times \mathbf{L})$, which is nothing but the orbital Rashba interaction induced by the polar displacement. Note that $\alpha_L \hat{\mathbf{z}} \cdot (\mathbf{k} \times \mathbf{L}) = i\alpha_L(k_+ L_- - k_- L_+)/2$ in terms of the raising and lowering operators of orbital angular momentum. Here, $k_\pm \equiv k_x \pm ik_y = ke^{\pm i\theta_\mathbf{k}}$ and $L_\pm \equiv L_x \pm iL_y$.

Now we diagonalize the mixing between $p_r$ and $p_z$ by orbital Rashba effect. To this end, we define another unitary transform as

$$U_2 = e^{-S_2}, \quad \text{(S8a)}$$

$$S_2 = \frac{ik\alpha_L(k)}{E_p + (a_1 - a_2)k^2}\begin{pmatrix} 0 & 0 & 1 & 0 \\ 0 & 0 & 0 & 0 \\ 1 & 0 & 0 & 0 \\ 0 & 0 & 0 & 0 \end{pmatrix}. \quad \text{(S8b)}$$

Then, the unitary transform $U = U_1 U_2$ diagonalizes the Hamiltonian.

$$U^\dagger H U = \begin{pmatrix} E_p + a_1 k^2 & 0 & 0 & 0 \\ 0 & E_p + a_2 k^2 & 0 & 0 \\ 0 & 0 & a_2 k^2 & 0 \\ 0 & 0 & 0 & E_s + bk^2 \end{pmatrix}, \quad \text{(S8c)}$$

and the perturbed eigenstates are $(U|p_r\rangle_{\text{Bi}}, U|p_t\rangle_{\text{Bi}}, U|p_z\rangle_{\text{Bi}}, U|s\rangle_{\text{Ag}})$.

To calculate the in-plane BC, the out-of-plane position operator $z$ is necessary. As demonstrated in Fig. S1(d), it consists of three contributions: the non-zero matrix elements between the nearest neighbour *s* and $p_z$ orbitals (denoted by $\zeta$), the nearest neighbour *s* and $p_r$ orbitals (denoted by $\eta$), and the on-site position difference of *s* orbital (denoted by $\Delta z$).

$$z = \begin{pmatrix} 0 & 0 & 0 & -i\eta k \\ 0 & 0 & 0 & 0 \\ 0 & 0 & 0 & \zeta \\ i\eta k & 0 & \zeta & -\Delta z \end{pmatrix}. \quad \text{(S9)}$$

Here the $\eta$ ($\zeta$) element is real because of its antisymmetric (symmetric) nature [Fig. S1(d)].

Note that $\eta$ is linearly proportional to $\Delta z$, but $\zeta$ is finite regardless of the polar displacement.

Now one can evaluate Eq. (2) to calculate the in-plane BC for the $p_z$ orbital, which is located at the Fermi level. The off-diagonal element of the velocity operator satisfies

$$\langle n'|\nabla_{\mathbf{k}} H|n\rangle = (E_n - E_{n'})\langle n'|\nabla_{\mathbf{k}}|n\rangle, \tag{S10a}$$

for eigenstates $n, n'$. For $|n\rangle = U|p_z\rangle$,

$$\langle n'|U^\dagger(\nabla_{\mathbf{k}} H)U|p_z\rangle = (a_2 k^2 - E_{n'})\langle n'|(U^\dagger \nabla_{\mathbf{k}} U)|p_z\rangle, \tag{S10b}$$

Now, the in-plane BC of the $p_z$ band is given by

$$\begin{aligned}\Omega_{\text{in}}^{p_z} &= -2\text{Re} \sum_{m \neq p_z} \frac{\langle p_z|U^\dagger(\hat{\mathbf{z}} \times \nabla_{\mathbf{k}} H)U|m\rangle\langle m|U^\dagger z U|p_z\rangle}{a_2 k^2 - E_m} \\ &= -2\text{Re} \sum_{m \neq p_z} \langle p_z|(\hat{\mathbf{z}} \times U^\dagger \nabla_{\mathbf{k}} U)|m\rangle\langle m|U^\dagger z U|p_z\rangle \\ &= \frac{4(a_2 - b)\zeta\xi}{[E_s - (a_2 - b)k^2]^2} \hat{\mathbf{z}} \times \mathbf{k},\end{aligned} \tag{S11}$$

which gives Eq. (4).

## V. Model for the in-plane BC in a GdAg$_2$ monolayer

To construct a model Hamiltonian, we resolve the spin and orbital characters of the bands (Fig. S2). Due to the large exchange splitting, we focus on spin-up bands [red lines in Fig. S2(a)] and develop a three-band Hamiltonian. When the majority-spin bands are denoted by $|1\rangle, |2\rangle$ and $|3\rangle$ states [Fig. S2(b)], $|1\rangle$ and $|3\rangle$ are given by $\mathbf{k}$-dependent linear combinations of $|d_{x^2-y^2}\rangle$ and $|d_{xy}\rangle$ and $|2\rangle$ is given by a $\mathbf{k}$-dependent mixture of $|d_{xz}\rangle$ and $|d_{yz}\rangle$ [Fig. S2(c)]. In an isotropic model, the basis kets are expressed as $|1\rangle = \cos\theta_{\mathbf{k}}|d_{x^2-y^2}\rangle + \sin\theta_{\mathbf{k}}|d_{xy}\rangle$, $|2\rangle = \cos\theta_{\mathbf{k}}|d_{xz}\rangle + \sin\theta_{\mathbf{k}}|d_{yz}\rangle$, and $|3\rangle = -\sin\theta_{\mathbf{k}}|d_{x^2-y^2}\rangle + \cos\theta_{\mathbf{k}}|d_{xy}\rangle$.

Within $(|1\rangle, |2\rangle, |3\rangle)$ basis set, the kinetic part of the Hamiltonian is modelled by

$$H_0 = \begin{pmatrix} E_1(\mathbf{k}) & 0 & 0 \\ 0 & E_2(\mathbf{k}) & 0 \\ 0 & 0 & E_3(\mathbf{k}) \end{pmatrix}, \tag{S12}$$

where $E_i$'s depend on $k$ only. In the presence of ferromagnetic moment aligned along the $x$ direction ($\mathbf{M} \parallel +\hat{\mathbf{x}}$), spin-orbit coupling mixes the orbital states with different angular momenta $m_z$. Under the strong exchange splitting, the orbital mixing occurs within the same spin channels and the spin-orbit coupling Hamiltonian ($H_1$) can be approximated as $\alpha \mathbf{L} \cdot$

$\mathbf{S} \sim \alpha \mathbf{L} \cdot \hat{x}$, where $\alpha$ is the spin-orbit coupling strength. The non-zero matrix elements of $L_x$ operator within the three $d$-orbital basis are $\langle d_{x^2-y^2}|L_x|d_{yz}\rangle = -\langle d_{xy}|L_x|d_{xz}\rangle = i$. As a result, the spin-orbit coupling term reads

$$H_1 = \alpha \begin{pmatrix} 0 & 0 & 0 \\ 0 & 0 & -i \\ 0 & i & 0 \end{pmatrix}. \tag{S13}$$

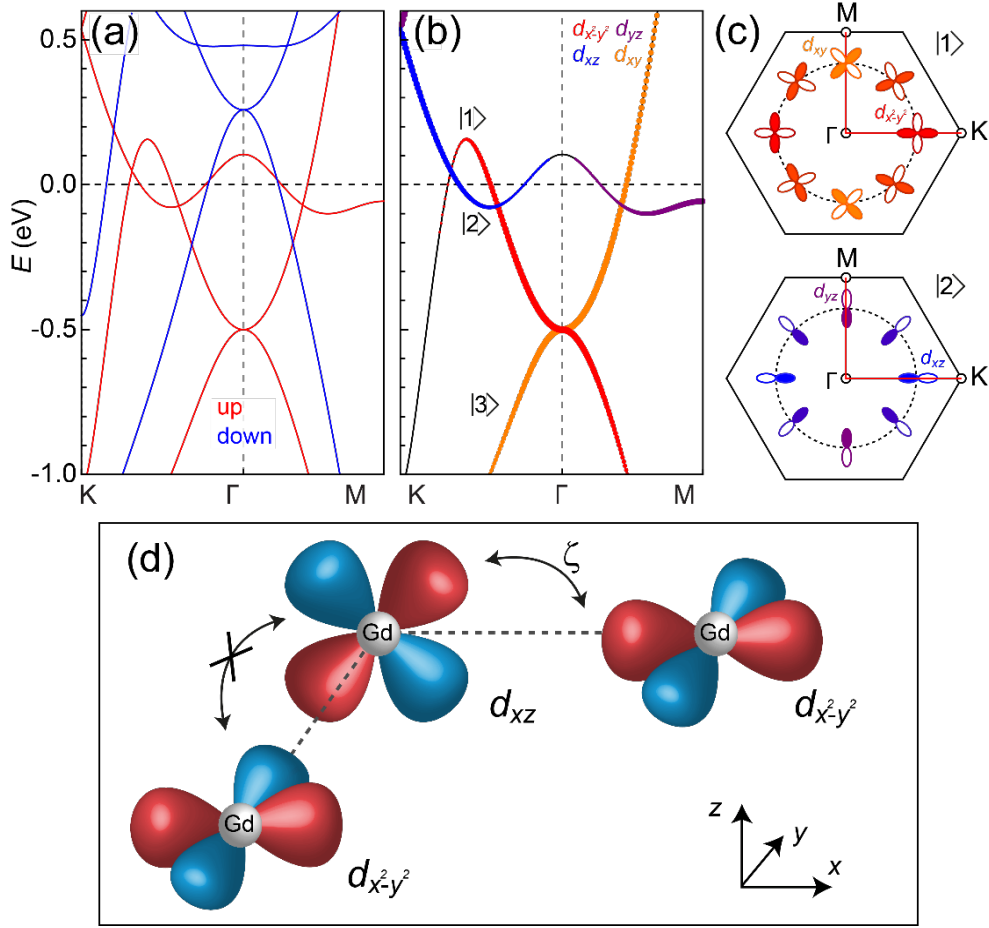

FIG. S2. (a) The spin character of a GdAg$_2$ monolayer. Spin-up and spin-down bands are presented by red and blue lines. (b) The orbital characteristics of spin-up electrons. The weights of $d_{x^2-y^2}$, $d_{xy}$, $d_{yz}$, and $d_{xz}$ are denoted by red, orange, purple, and blue dots. (c) The orbital textures in the momentum space for $|1\rangle$ and $|2\rangle$. The orbital texture for $|3\rangle$ is given by 90-degree rotation of that for $|1\rangle$ in the momentum space. (d) Illustration of a non-zero position matrix element between $d_{xz}$ and $d_{x^2-y^2}$ orbitals depending on the hopping direction.

To diagonalize the total Hamiltonian $H = H_0 + H_1$, we treat $\alpha$ pertubatively and keep its first order contributions. After some algebra, keeping first order contributions in $\alpha$, we may diagonalize the total Hamiltonian by

$$e^S(H_0 + H_1)e^{-S} = \begin{pmatrix} E_1 & 0 & 0 \\ 0 & E_2 & 0 \\ 0 & 0 & E_3 \end{pmatrix}, \tag{S14a}$$

$$S = -\frac{i\alpha}{E_2 - E_3}\begin{pmatrix} 0 & 0 & 0 \\ 0 & 0 & 1 \\ 0 & 1 & 0 \end{pmatrix}. \tag{S14b}$$

The perturbed eigenstates are then $(e^{-S}|1\rangle, e^{-S}|2\rangle, e^{-S}|3\rangle)$.

For the velocity off-diagonal component, we use Eq. (S10a). For simplicity, we may approximate $S$ to be independent of $\mathbf{k}$, $\langle n'|e^S(\nabla_\mathbf{k} e^{-S}|n\rangle) \approx \langle n'|\nabla_\mathbf{k}|n\rangle$, whose error is higher order in $1/(E_2 - E_3)$. Then the BC for $|1\rangle$, which is close to the Fermi level, is then straightforwardly calculated by Eq. (2).

$$\mathbf{\Omega}_{\text{in}}^{|1\rangle} = 2\hat{\mathbf{z}} \times \text{Re}[\langle 1|\nabla_\mathbf{k}|3\rangle\langle 3|e^S z e^{-S}|1\rangle]. \tag{S15}$$

By using $|1\rangle = \cos\theta_\mathbf{k} |d_{x^2-y^2}\rangle + \sin\theta_\mathbf{k} |d_{xy}\rangle$ and $|3\rangle = -\sin\theta_\mathbf{k} |d_{x^2-y^2}\rangle + \cos\theta_\mathbf{k} |d_{xy}\rangle$, we obtain

$$\langle 1|\nabla_\mathbf{k}|3\rangle = -\langle 3|\nabla_\mathbf{k}|1\rangle = \frac{\mathbf{k} \times \hat{\mathbf{z}}}{k^2}. \tag{S16}$$

For the position operator, we illustrate the matrix element of $z$ between $d_{x^2-y^2}$ and $d_{xz}$ orbitals in Fig. S2(d), which implies

$$\langle d_{x^2-y^2}|z|d_{xz}\rangle = i\zeta k_x \quad (k_x \text{ direction}), \tag{S17a}$$

$$\langle d_{x^2-y^2}|z|d_{xz}\rangle = 0 \quad (k_y \text{ direction}). \tag{S17b}$$

In our isotropic model, Eqs. (S17a)-(S17b) can be expressed as

$$\langle d_{r^2-t^2}|z|d_{rz}\rangle = i\zeta k, \tag{S17c}$$

$$\langle d_{r^2-t^2}|z|d_{tz}\rangle = 0, \tag{S17d}$$

where $r = \cos\theta_\mathbf{k} x + \sin\theta_\mathbf{k} y$ ($t = -\sin\theta_\mathbf{k} x + \cos\theta_\mathbf{k} y$) denotes the radial (tangential) direction. Note that the $d$ orbitals in the $(r, t)$ coordinate can be expressed by those in the $(x, y)$ coordinate by the following rotational transform.

$$\begin{pmatrix} |d_{x^2-y^2}\rangle \\ |d_{xy}\rangle \\ |d_{yz}\rangle \\ |d_{xz}\rangle \end{pmatrix} = \begin{pmatrix} \cos 2\theta_\mathbf{k} & -\sin 2\theta_\mathbf{k} & 0 & 0 \\ \sin 2\theta_\mathbf{k} & \cos 2\theta_\mathbf{k} & 0 & 0 \\ 0 & 0 & \cos\theta_\mathbf{k} & \sin\theta_\mathbf{k} \\ 0 & 0 & -\sin\theta_\mathbf{k} & \cos\theta_\mathbf{k} \end{pmatrix}\begin{pmatrix} |d_{r^2-t^2}\rangle \\ |d_{rt}\rangle \\ |d_{tz}\rangle \\ |d_{rz}\rangle \end{pmatrix}. \tag{S18}$$

By using Eqs. (S17c)-(S18) and the definitions of $|1\rangle, |2\rangle, |3\rangle$, one can straightforwardly show that the non-zero matrix elements of $z$ are given by

$$z = i\zeta \begin{pmatrix} 0 & k_x & 0 \\ -k_x & 0 & -k_y \\ 0 & k_y & 0 \end{pmatrix}. \tag{S19}$$

As a remark, there is another non-zero matrix element of $z$ (not shown in Fig. S2), $\langle d_{rt}|z|d_{tz}\rangle = i\zeta' k$ and $\langle d_{rt}|z|d_{rz}\rangle = 0$, but its contributions are cancelled by the rotational transform, Eq. (S18). Also, the other elements of $z$ are zero. Therefore, Eq. (S17c) is the only hopping that contributes to $z$. From Eq. (S19), a straightforward algebra gives the following transformed $z$ matrix.

$$e^S z e^{-S} = z + \frac{\alpha\zeta}{E_2 - E_3} \begin{pmatrix} 0 & 0 & k_x \\ 0 & -2k_y & 0 \\ k_x & 0 & 2k_y \end{pmatrix}. \tag{S20}$$

Plugging Eqs. (S16) and (S20) into Eq. (S15) gives the in-plane BC as

$$\Omega_{\text{in}}^{|1\rangle} = \frac{2\alpha\zeta k_x}{k^2(E_2 - E_3)} \mathbf{k}, \tag{S21}$$

which gives Eq. (5).

## VI. Details for time-dependent DFT calculations

To investigate the anomalous Hall current induced by the in-plane BC, we conducted time-dependent DFT [39] calculations using the real-time propagation package implemented in the ELK code. The wave function was evolved through the time-dependent Kohn–Sham Hamiltonian as

$$i\frac{\partial\psi_{n\mathbf{k}}(\mathbf{r},t)}{\partial t} = \left\{\frac{1}{2}\left|-i\boldsymbol{\nabla} + \frac{1}{c}\mathbf{A}_{\text{ext}}(t)\right|^2 + v_{\text{eff}}(\mathbf{r},t)\right\}\psi_{n\mathbf{k}}(\mathbf{r},t), \tag{S22}$$

where $n$ is the band index and $\mathbf{k}$ denotes the wave vector in the Brillouin zone. In our calculations, the external field is expressed by the velocity gauge of the vector potential $\mathbf{E}_{\text{ext}}(t) = -(1/c)\partial\mathbf{A}_{\text{ext}}(t)/\partial t$. The external vector potential $\mathbf{A}_{\text{ext}}(t)$ adopted in our calculation is expressed in the form $A\sin(\omega t)\exp[-(t-t_0)^2/2\sigma^2]\hat{\mathbf{x}}$. We used a short light pulse with an amplitude ($A$) of 10 a.u. and standard deviation ($\sigma$) of 0.32 fs, an envelope peak time ($t_0$) of 1.21 fs, and a photon energy ($\hbar\omega$) of 2.711 eV. According to the Courant-Friedrichs-Lewy convergence condition, the time-interval $dt$ for the wave function propagation is bound by the fastest motion in the system. In our case, it occurs at the confined motion of electrons along the $z$ direction shown as high-frequency small oscillations around $t = 1$ fs [see Figs. 4(b)-(c)]. We used $dt = 0.1$ a.u. to guarantee the validity of long-time dynamics.

## VII. Details for normalization of the angle-dependent Hall currents

In the time-dependent DFT calculations for $J_z(\theta, t)$ in Fig. 4(c), we have seven data sets for $\theta = 0°, 30°, 60°, 90°, 120°, 150°, 180°$. If the current originates from the linear anomalous Hall effect from the in-plane BC as our theory implies, the data are expected to satisfy $J_z(\theta, t) = \mathcal{J}_0(t) \cos\theta$ for a normalization factor $\mathcal{J}_0(t)$ at each evaluation time $t$. Although the normalization factor is dependent on time, the cosine function of the angle dependence should be universal. Therefore, we find the proper normalization factor $\mathcal{J}_0(t)$ and compare the normalized Hall current $J_z(\theta, t)/\mathcal{J}_0(t)$ with $\cos\theta$.

Although $\mathcal{J}_0(t) = J_z(0°, t)$ appears to give a simple relation between our numerical data and the normalization factor, it yields divergence because $J_z(0°, t) \approx 0$ at several evaluation times. Instead, when the mathematical identity $|\mathcal{J}_0(t)| = \frac{1}{2}\sqrt{\sum_{\{\theta\}}[\mathcal{J}_0(t)\cos\theta]^2}$ is used, the appropriate normalization factor is given by $\frac{1}{2}\sqrt{\sum_{\{\theta\}}[J_z(\theta, t)]^2}$, where the summation runs over the seven angles $\{\theta\} = \{0°, 30°, 60°, 90°, 120°, 150°, 180°\}$. This approach does not result in divergence unless the data are zero for all angles, which is a trivial case. Because the sign of $\mathcal{J}_0(t)$ is the same as that of $J_z(0°, t)$, we define the following normalized current:

$$\tilde{J}_z(\theta, t) = \frac{2 J_z(\theta, t)}{\text{sgn}[J_z(0°, t)] \sqrt{\sum_{\{\theta\}}[J_z(\theta, t)]^2}}. \tag{S23}$$

The right-hand side can now be evaluated solely from the simulation data. If our theory is valid, this value should universally follow $\cos\theta$ irrespective of the value of $t$.

We calculate $\tilde{J}_z(\theta, t)$ over *all* the evaluation times after $t \geq 0.86$ fs, the number of which is 382 for each of the angles and is thus 2674 in total. For $t < 0.86$ fs, $J_z(\theta, t)$ is too small for the cosine dependence of the normalized currents to be reasonably observed. In Fig. 4(e), the mean values of $\tilde{J}_z(\theta, t)$ are plotted as orange dots and the standard deviations as red error bars; the data match perfectly with the expected cosine function [solid guideline in Fig. 4(e)].